# On the bosonic atoms

M. Ya. Amusia[1, 21] and L. V. Chernysheva[2]

[1]*The Racah Institute of Physics, the Hebrew University, Jerusalem 91904, Israel*
[2]*A. F. Ioffe Physical-Technical Institute, St. Petersburg 194021, Russian Federation*

**Abstract:**
We investigate here ground state properties of atoms, in which substitute fermions – electrons by bosons, e.g. $\pi^-$ - mesons. We perform some calculations in the frame of modified Hartree-Fock (HF) equation.

The modification takes into account the symmetry, instead of anti-symmetry of the two-identical bosons wave function. The modified HF approach thus enhances (doubles) the effect of self-action for the boson case. Therefore, we accordingly modify the HF equations by eliminating from them the self-action terms "by hand".

The contribution of meson-meson and meson-nucleon non-Coulomb interaction is inessential at least for atoms with low and intermediate nuclear charge, which is our main aim.

We found the binding energy of pion negative ions $A_\pi^-$ of pion atoms $A_\pi$ and the number of extra bound pions $\Delta N_\pi$ increases with the growth of nuclear charge Z. For e.g. Xe $\Delta N_\pi = 4$.

As an example of a simple process with a pion atom, we consider photoionization that differs essentially from that for electron atoms. Namely, it is not monotonic decreasing from the threshold but has a prominent maximum above threshold instead. We study also elastic scattering of pions by pionic atoms.

**1**. Heavy ion colliders are able to create very big numbers of $\pi^\mp$ - mesons, or pions in each process of ion collision. The projectile or target ions can capture negative pions during the nuclear interaction process and form so-called pionic atoms. The formation of these atoms becomes a source of photon emission with quite specific spectrum, the shape of which is determined by the lowest possible binding energy of a pion in the nuclear field. Each ion or atom can capture several pions so the interaction between them can be essential and one has to consider this interaction. The structure of such an object is interesting by itself. This is the aim of this Letter.

Currently, the starting point in investigating ground state properties and processes in ordinary (or electronic) atoms are the Hartree-Fock (HF) equations. It is natural to apply the properly adjusted HF equations also to pionic atoms. However, it proved to be not straightforward. Thus, the second aim of this Letter is the demonstration of the unexpected feature of these equations.

**2.** The HF equation for multi-electron atoms looks like (see e.g. [1])[2]

---

[1]amusia@vms.huji.ac.il

$$-\frac{\Delta}{2}\phi_j(x)-\frac{Z}{r}\phi_j(x)+\sum_{k=1}^{N_e}\int \phi_k^*(x')\frac{dx'}{|\mathbf{r}-\mathbf{r}'|}\left[\phi_k(x')\phi_j(x)-\phi_j(x')\phi_k(x)\right]=E_j\phi_j(x). \qquad (1)$$

Here Z is the nuclear charge, $\phi_j(x)$ is the one electron wave function, $x \equiv \vec{r},\vec{\sigma}$ are the combination of electron coordinate and spin variables, $E_j$ - is the one-electron or so-called HF energy, the summation is performed over all occupied electron states $N_e$.

The second term in the square brackets in (1) is named Fock term. The sign "–"in front of it is a trace of the Fermi-nature of electrons. It automatically removes the self-action, i.e. the action of an atomic electron upon itself. Indeed, because of its action the contribution from the term $k = j$ in the sum over $k$ in (1) disappear.

An equation similar to (1) for bosons would be

$$-\frac{\Delta}{2}\chi_j(x)-\frac{Z}{r}\chi_j(x)+\sum_{k=1}^{N_\pi}\int \chi_k^*(x')\frac{dx'}{|\mathbf{r}-\mathbf{r}'|}\left[\chi_k(x')\chi_j(x)+\chi_j(x')\chi_k(x)\right]=E_j\chi_j(x). \qquad (2)$$

Here"+" instead of "-"in the square bracket takes into account that identical bosons wave function is symmetric against permutation of the coordinate/spin variables. As a result, this equation for bosons does not eliminate self-action: it only enhances self-action. The summation in (2) goes over all pion states $N_\pi$.

If the interparticle interaction is not long range Coulomb $|\mathbf{r}-\mathbf{r}'|^{-1}$, but short-range $u_0\delta(|\mathbf{r}-\mathbf{r}'|)$, and the interaction that holds the system together is not the nuclear Coulomb potential $-Z/r$, but something more general $U(r)$, equation (2) reduces to so-called Gross-Pitaevskii equation [2]

$$-\frac{\Delta}{2}\chi_j(x)+U(r)\chi_j(x)+u_0\sum_{k=1}^{N_\pi}|\chi_k(r)|^2\chi_j(r)=E_j\chi_j(r). \qquad (3)$$

Eq. (3) includes self-action that is acceptable for big system, but since we are interested in light and medium atoms, we have to use a version of (2) that is obtained by eliminating in the sum over $k$ the term $k = j$. Since all atomic pions can be in one lowest state $k$, we have to solve the following equation

$$-\frac{\Delta}{2}\chi_j(x)-\frac{Z}{r}\chi_j(x)+(N_\pi-1)\int \chi_k^*(x')\frac{dx'}{|\mathbf{r}-\mathbf{r}'|}\left[\chi_k(x')\chi_j(x)+\chi_j(x')\chi_k(x)\right]=E_j\chi_j(x). \qquad (4)$$

---

[2] We employ here are system of units that defines $m = e = \hbar = 1$. Here $m$ is the mass, in electron case $m = m_e$, in pion case $m = m_\pi$; $m_\pi/m_e \approx 273.13$.

Assuming all pions occupying the same level, on has instead of (4) the equation

$$-\frac{\Delta}{2}\chi_k(\mathbf{r}) - \frac{Z}{r}\chi_k(\mathbf{r}) + 2(N_\pi^k - 1)\int |\chi_k(\mathbf{r})|^2 \frac{d\mathbf{r}'}{|\mathbf{r}-\mathbf{r}'|} = E_k \chi_k(\mathbf{r}). \qquad (5)$$

Here the relation $\rho_k(\mathbf{r}) = N_{\pi,k} |\chi_k(\mathbf{r})|^2$ determines the pion density $\rho_k(r)$.

**3.** We have solved Eq (4) numerically and obtained HF energies of the 1s level, total energies and mean square radiuses for pionic atoms and negative pionic ions with Z from 2 to 80. Table 1 presents the results for a number of pionic atoms and negative ions for He, Ne, Ar, Co, Kr and Xe, and ordinary noble gas atoms.

**Table 1.** 1s level $E_{1s}$, ground state total energy $E_{tot}$, mean square radius $\langle r^2 \rangle$ of $\pi$-meson atoms, and negative ions along with ordinary atoms

| Pionic atoms | $-E_{1s}$ | $-E_{tot}$ | $\langle r^2 \rangle$ | Ordinary atoms | $-E_{1s}$ | $-E_{tot}$ | $\langle r^2 \rangle$ |
|---|---|---|---|---|---|---|---|
| He 1s$^2$ | 1.8359 | 5.723 | 1.1847 | He 1s$^2$ | 1.8359 | 5.723 | 1.1847 |
| He 1s$^3$ | 0.3679 | 5.856 | 2.3036 | | | | |
| He 1s$^4$ | no | | | | | | |
| Ne 1s$^{10}$ | 15.083 | 528.70 | 0.07912 | Ne 1s$^2$ | 65.545 | 257.09 | 0.0335 |
| Ne 1s$^{11}$ | 9.244 | 536.72 | 0.0941 | | | | |
| Ne 1s$^{14}$ | no | | | | | | |
| Ar 1s$^{18}$ | 40.14 | 2976.96 | 0.0263 | Ar 1s$^2$ | 237.22 | 1053.64 | 0.0099 |
| Ar 1s$^{19}$ | 29.95 | 3003.69 | 0.0290 | | | | |
| Ar 1s$^{20}$ | no | | | | | | |
| Co 1s$^{27}$ | 82.47 | 9895.57 | 0.0121 | | | | |
| Co 1s$^{28}$ | 67.39 | 9959.63 | 0.0129 | | | | |
| Co 1s$^{29}$ | 53.15 | 10009.3 | 0.0139 | | | | |
| Co 1s$^{30}$ | no | | | | | | |
| Kr 1s$^{36}$ | 139.78 | 23281.9 | 0.00689 | Kr 1s$^2$ | 1040.33 | 5504.11 | 0.0024 |
| Kr 1s$^{37}$ | 119.81 | 23397.2 | 0.00726 | | | | |
| Kr 1s$^{38}$ | 100.67 | 23493.3 | 0.00767 | | | | |
| Kr 1s$^{39}$ | 82.40 | 23570.9 | 0.00812 | | | | |
| Kr 1s$^{40}$ | no | | | | | | |
| Xe 1s$^{54}$ | 299.32 | 77991.3 | 0.00312 | Xe 1s$^2$ | 2448.80 | 14464.3 | 0.00106 |
| Xe 1s$^{55}$ | 269.57 | 78254.2 | 0.00323 | | | | |
| Xe 1s$^{56}$ | 240.66 | 78488.0 | 0.00335 | | | | |
| Xe 1s$^{57}$ | 212.60 | 78693.6 | 0.00347 | | | | |
| Xe 1s$^{58}$ | 185.41 | 78871.8 | 0.00361 | | | | |
| Xe 1s$^{59}$ | no | | | | | | |

According to the data presented in Table 1, the total binding energy of a pion atom is much bigger than that for an ordinary atom, while the size of the pion atom is several times bigger than the size of a normal atom. Sizes and energies we express in pionic and normal atomic units, respectively. The ionization potential (-$E_{1s_\pi}$) of a pionic atom is much smaller than that of a neutral atom.

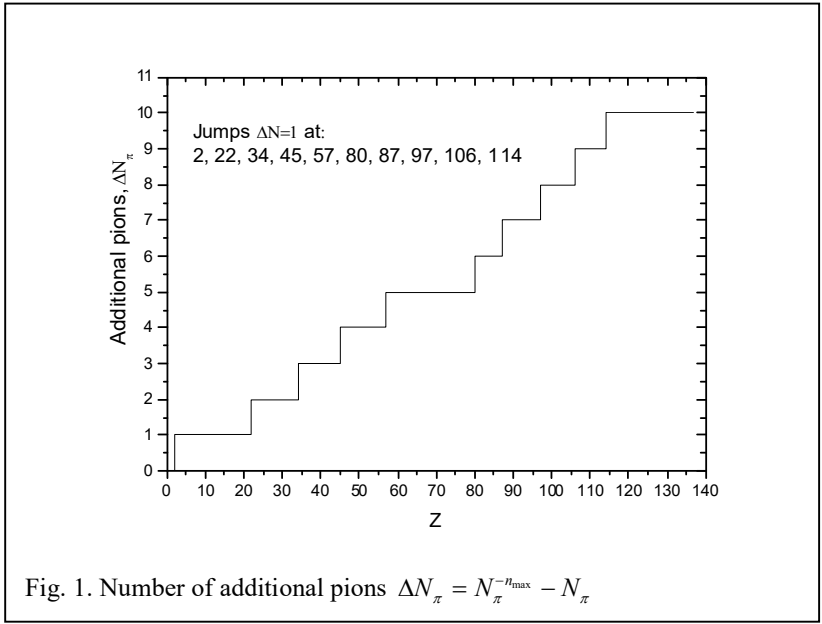

Fig. 1. Number of additional pions $\Delta N_\pi = N_\pi^{-n_{\max}} - N_\pi$

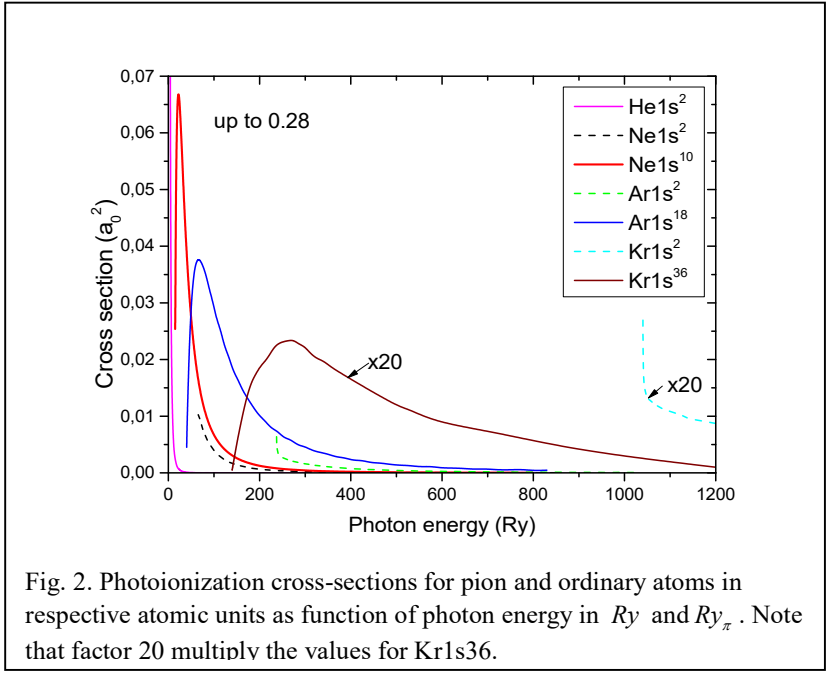

Fig. 2. Photoionization cross-sections for pion and ordinary atoms in respective atomic units as function of photon energy in $Ry$ and $Ry_\pi$. Note that factor 20 multiply the values for Kr1s36.

An interesting feature is the ability to form easily negative ions. For pionic He, Ne and Ar we predict the possible formation of single negative pionic ions $He_\pi^-, Ne_\pi^-, Ar_\pi^-$. For $Co$, the formation of $Co_\pi^-$ and $Co_\pi^{-2}$ becomes possible. $Kr$ is able to form up to $Kr_\pi^{-3}$, while $Xe$ is able to attach extra four pions that leads to $Xe_\pi^{-4}$. We have extended calculations up to Z=137. At this value, the Coulomb non-relativistic binding energy is close to the rest mass of a pion. Note that for pion atoms such a notion as "noble gas" does not exist, since pion shells does not exist also. Note also that such big values of Z, and even bigger are achieved temporarily, during fast heavy ion collisions.

Fig. 1 presents the dependence of the number of extra bound pions $\Delta N_\pi = N_\pi^{-n_{\max}} - N_\pi$ as a function of nuclear charge Z. The biggest steps in variation of Z needed to reach $\Delta N_\pi = 1$ is 2–22 and 57–80. The first half of the latter coincide with lanthanides that seems to be accidental.

**4.** Of some interest is to study the processes of interaction of projectiles with pion atoms. Let us consider at first photoionization. Eq. 5 for $E \geq 0$ determines the continuous spectrum wave functions $\chi_E(\mathbf{r})$, which is automatic orthogonal to the wave function of the occupied level. The following relation determines the cross-section (see e.g. [3]):

$$\sigma_k(\omega) = \frac{4\pi^2 N_\pi}{\omega c} \left| \langle \chi_{\omega+E_k} | \hat{d} | \chi_k \rangle \right|^2. \qquad (6)$$

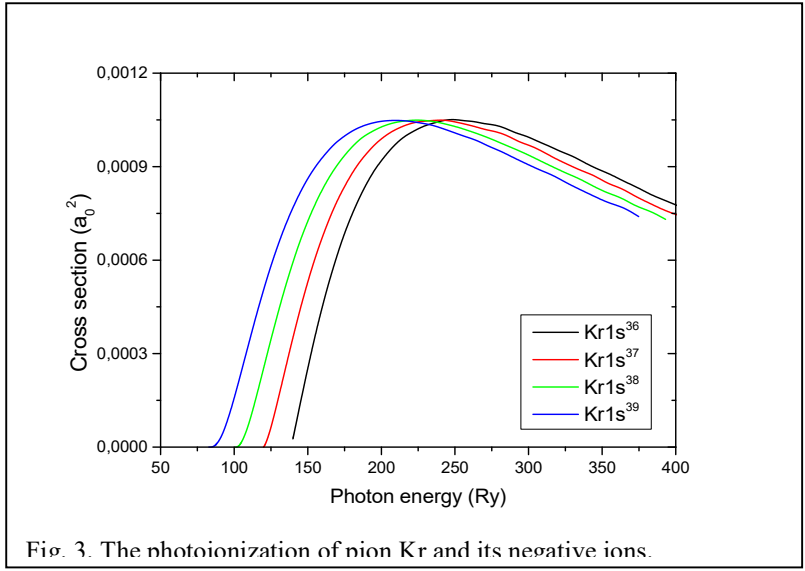

Fig. 3. The photoionization of pion Kr and its negative ions.

Here $\hat{d}$ is the photon-pion interaction operator; $c$ is the speed of light.

Fig. 2 presents the results of calculations using for $\sigma_k(\omega)$ the computing code ATOM-M [4] for He, Ne, Kr and Ar. We compare the results for pionic atoms to that for ordinary atoms, presented in respective atomic units.

For ordinary atoms, the cross-section monotonically decreases with photon energy growth from the maximum value at threshold. For pion atom the cross-section, on the contrary, reaches its maximum value above ionization threshold, thus forming pronounce maxima, the height of which is decreasing while the width is increasing with Z growth. The area under photoionization curve satisfies, quite naturally, the sum rule (see e.g. [3]).

Fig. 3 presents the dependence of the photoionization cross-section upon $\Delta N$ for $\Delta N = 0 \div 3$. We see that with increase of $\Delta N$ the ionization threshold decreases, and the cross-section almost as a whole move to lower energies. Note that at threshold the pion cross-section has a typical for long-range Coulomb field, but small compared to ordinary atoms, jump for neutral atoms. For ions, the cross section starts to increases from zero at threshold.

Note that we were unable to find discrete excitations in pion negative ions, contrary to the case of neutrals.

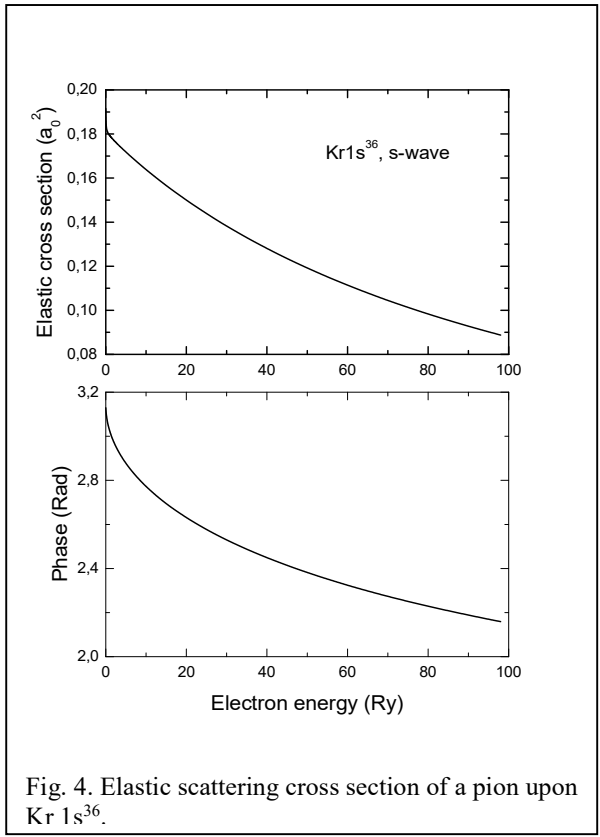

Fig. 4. Elastic scattering cross section of a pion upon Kr $1s^{36}$.

Fig. 4 presents the cross section and s-phase of elastic scattering of pions upon Kr 1s$^{36}$. All $l > 0$ phases are negligibly small in the considered energy range. The s-phase riches $\pi$ at zero energy, thus signaling the presence of a single charged negative ion Kr 1s$^{37}$. From Table 1 we see that in fact even a triply charged Kr 1s$^{39}$ exists. It means that formation of doubly and triply charged negative ions takes place due to self-consistent variation of the field of a negative ion with growth of N. Note that in investigation of a $\pi^+$ scattering upon pionic atom, e.g. Kr 1s$^{36}$, the interaction of $\pi^+$ and $\pi^-$ with formation of a bound pionium could be very important.

Electron cross-section scattering upon pion atom, e.g. the considered above Kr 1s$^{36}$, is by a factor $\beta^2 \equiv (m_e/m_\pi)^2 \simeq 1.34 \times 10^{-5}$, i.e. negligibly small [5].

**5.** We treated, for the first time, a purely multi-pionic atom, their properties and processes with their participation. The results obtained are to some extent surprising and considerably different from that for ordinary atoms. In the consideration presented above, we neglected the nuclear forces between pions and nucleons. They can inforce sticking of extra pions to a nucleus but become essentially important only at $Z \geq 100$.

It would be of interest to look for data in multi-pion atoms formed in heavy atom collisions. It would be of interest also to go in studies of a boson atom beyond the self-consistent field approximation. Note that S. T. Belyaev and a number of other people have calculated the energy and excitation spectrum for an infinite homogeneous Bose-gas with short-range inter-boson interaction already long ago (see e.g. [6]).